\documentclass[10pt,a4paper,twocolumn]{article}

\usepackage[utf8]{inputenc}
\usepackage[T1]{fontenc}
\usepackage{amsmath,amssymb,amsthm}
\usepackage{mathtools}
\usepackage{booktabs}
\usepackage{hyperref}
\usepackage[margin=1.8cm,top=2cm,bottom=2cm]{geometry}
\usepackage{enumitem}
\usepackage{tikz}
\usetikzlibrary{arrows.meta,positioning,calc,shapes.geometric,fit,backgrounds}
\usepackage{xcolor}
\usepackage{multirow}
\usepackage{caption}
\usepackage{tabularx}
\usepackage{url}
\usepackage{graphicx}
\usepackage{float}
\usepackage{array}
\usepackage{pgfplots}
\pgfplotsset{compat=1.18}
\usepackage{tcolorbox}
\tcbuselibrary{breakable}

\definecolor{trusthigh}{HTML}{2E86C1}
\definecolor{trustmid}{HTML}{F39C12}
\definecolor{trustlow}{HTML}{E74C3C}
\definecolor{trustrecov}{HTML}{27AE60}
\definecolor{siloA}{HTML}{3498DB}
\definecolor{siloB}{HTML}{E67E22}
\definecolor{siloC}{HTML}{2ECC71}
\definecolor{examplebox}{HTML}{EBF5FB}

\newtheorem{definition}{Definition}[section]

\newcommand{\TpW}{\mathrm{TpW}}

\title{\textbf{Zero Trust for Multi-RAT IoT: \\Trust Boundary Management in \\Heterogeneous Wireless Network Environments}}

\author{
Jonathan Shelby \\
Department of Computer Science, University of Oxford\\
\texttt{jonathan.shelby@cs.ox.ac.uk}
}
\date{}

\begin{document}

\maketitle

\begin{abstract}
The proliferation of Multi-Radio Access Technology (multi-RAT) Internet of Things devices---particularly Unmanned Aerial Vehicles (UAVs) operating across LoRaWAN, 5G/4G cellular, Meshtastic mesh, proprietary protocols such as DJI OcuSync, MAVLink telemetry links, Wi-Fi, and satellite---creates a fundamental and hitherto unexamined challenge for Zero Trust Architecture (ZTA) adoption. Each transition between radio access technologies constitutes a \textit{trust boundary crossing}: the device exits one network trust domain and enters another, potentially invalidating authentication state, device attestation, and contextual trust signals. Current ZTA frameworks assume relatively stable network environments and do not address the trust implications of frequent, dynamic RAT switching in mobile IoT deployments.

We present a systematic qualitative analysis of trust-state management in multi-RAT IoT environments, with UAV operations as a primary case study. We formalise the concept of a \textit{trust boundary crossing event}, characterise trust state transitions across seven representative RAT families---including both open-standard and proprietary drone communication protocols---and introduce \textit{trust portability} as a first-class architectural concept: the capacity for cryptographic trust evidence to survive RAT handover at bounded computational and energy cost. We identify a fundamental tension between Zero Trust's ``never trust, always verify'' mandate and the resource constraints of mobile IoT devices that must re-establish trust at every network transition. Our analysis demonstrates that na\"ive ZTA implementation in multi-RAT environments incurs re-authentication costs that can consume a significant fraction of a constrained device's communications power budget, motivating the need for power-aware trust optimisation metrics. We develop a four-category transition taxonomy (planned, coverage-driven, opportunistic, adversary-forced) with distinct trust cost multipliers, introduce temporal trust decay modelling, extend the RAT analysis to eight families including Bluetooth Low Energy, define formal security properties for trust portability artefacts, analyse trust composition for parallel multi-RAT operation, and contextualise the framework within EU and FAA Remote ID regulations. We conclude by identifying formal requirements for power-aware trust metrics, establishing the foundation for subsequent work on trust-efficient Zero Trust implementations in resource-constrained environments.
\end{abstract}


%=====================================================================
\section{Introduction}
\label{sec:intro}
%=====================================================================

Zero Trust Architecture (ZTA) has emerged as the dominant security paradigm for modern networked systems, driven by the recognition that perimeter-based defences are insufficient against increasingly sophisticated threats~\cite{nist800207,kindervag2010}. The core principle---``never trust, always verify''---mandates continuous authentication, authorisation, and validation of every device, user, and data flow regardless of network location. Adoption is accelerating: a 2024 Gartner survey found that 63\% of organisations worldwide have either partially or fully implemented a Zero Trust strategy~\cite{gartner2024zt}, and the United States Department of Defense has mandated ZTA adoption across all services by 2027~\cite{dodztref2021}.

Simultaneously, the Internet of Things (IoT) landscape is undergoing a parallel transformation. Mobile IoT devices---particularly Unmanned Aerial Vehicles (UAVs)---are shifting from single-radio to multi-RAT architectures that combine cellular (5G, 4G LTE), Low-Power Wide-Area Network (LPWAN) technologies (LoRaWAN, NB-IoT), mesh protocols (Meshtastic), proprietary links (DJI OcuSync, Lightbridge), open autopilot telemetry (MAVLink over SiK/RFD900 radios), Wi-Fi, and satellite connectivity~\cite{leenders2021multirat,pmc2023lorafanet}. This multi-RAT approach enables broader coverage, greater resilience, and cost-optimised connectivity across heterogeneous operational environments.

These two trends are on a collision course. ZTA demands continuous verification---but multi-RAT switching demands frequent \textit{re}-verification across fundamentally different trust domains. Each radio access technology embeds its own identity model, authentication mechanism, channel encryption scheme, and trust anchor. A device authenticated via 5G-AKA with a USIM-bound identity carries trust state that is meaningless to a LoRaWAN network server expecting OTAA join credentials, and entirely alien to a Meshtastic mesh operating on pre-shared keys. Proprietary protocols such as DJI's OcuSync present an additional dimension: their authentication mechanisms are opaque, undocumented, and architecturally incompatible with any external trust framework.

Despite the practical urgency of this intersection, the literature has not systematically examined what happens to Zero Trust compliance when a device crosses from one RAT to another. This paper addresses that gap.

\subsection{Motivating Scenario}

Consider a commercial UAV conducting an infrastructure inspection mission in rural terrain. The drone launches from an operations base connected via a private 5G cell, with full Zero Trust authentication: device identity verified via eSIM, firmware attestation via a Trusted Platform Module (TPM), behavioural baseline established, and policy enforcement active. As the drone flies beyond 5G coverage, it falls back to 4G LTE---a handover within the 3GPP family that preserves some trust state via the cellular authentication infrastructure, but triggers policy re-evaluation due to reduced Quality of Service.

The drone continues into an area with no cellular coverage and switches to LoRaWAN via a ground-deployed gateway. The LoRaWAN network uses an entirely different authentication model (Over-the-Air Activation with AES-128 key derivation), has no concept of device attestation, and offers no mechanism to receive or validate the trust evidence accumulated on the cellular network. From a Zero Trust perspective, the device's trust score drops precipitously.

During the survey, the drone encounters a neighbouring operator's Meshtastic mesh network, offering potential data relay via AES-256 pre-shared key encryption---but with no identity infrastructure, no mutual authentication, and no policy enforcement capability. Meanwhile, the drone's primary command-and-control link operates over a proprietary protocol (e.g., DJI OcuSync with AES-256 session encryption), running in parallel but architecturally isolated from the data network's trust model. The telemetry stream to a ground monitoring station uses MAVLink over an RFD900 radio---a protocol that, in its most widely deployed configuration, provides only optional HMAC-based message signing with no encryption~\cite{mavsec2019,mavlinkcve2020}.

On return, the drone re-enters cellular coverage and must re-establish full Zero Trust compliance on the 5G network. At each transition, the ZTA policy engine faces critical questions: What trust state survives the handover? What must be re-verified, and at what computational and energy cost? What is the minimum trust threshold for continued operation on the new RAT? Can trust evidence from the previous domain be presented to the new domain?

\subsection{Problem Statement}

Given a mobile IoT device $d$ traversing an ordered sequence of RATs $\langle R_1, R_2, \ldots, R_n \rangle$ under ZTA policy $\Pi$, how should the device manage trust state transitions at each boundary crossing $R_i \rightarrow R_{i+1}$ to maintain minimum trust thresholds $\mathcal{T}_{\min}$ while minimising cumulative re-authentication cost $C_{\mathrm{auth}}$ under power budget $P_{\max}$?

\subsection{Contributions}

This paper makes the following contributions:

\begin{enumerate}[nosep]
    \item A systematic characterisation of trust-relevant properties across eight IoT RAT families, including open-standard, proprietary, and short-range protocols (Section~\ref{sec:background}).
    \item A taxonomy of RAT transition types---planned, coverage-driven, opportunistic, and adversary-forced---with distinct trust cost implications (Section~\ref{sec:model}).
    \item A formalisation of \textit{trust boundary crossing events} with temporal trust decay modelling (Section~\ref{sec:model}).
    \item The introduction of \textit{trust portability} with formally specified security properties, and analysis of trust composition for parallel multi-RAT operation (Section~\ref{sec:portability}).
    \item A qualitative case study with a fully worked numerical example, analysing trust dynamics, power costs, and threat exposure (Section~\ref{sec:casestudy}).
    \item Contextualisation within the Remote ID regulatory landscape and identification of formal requirements for power-aware trust optimisation metrics (Section~\ref{sec:discussion}).
\end{enumerate}

\subsection{Paper Organisation}

Section~\ref{sec:background} provides background on ZTA and multi-RAT IoT architectures, with particular attention to drone-specific communication protocols. Section~\ref{sec:related} surveys related work and identifies the research gap. Section~\ref{sec:model} presents the trust boundary model. Section~\ref{sec:portability} introduces trust portability mechanisms. Section~\ref{sec:casestudy} develops the UAV case study. Section~\ref{sec:discussion} discusses implications and bridges to future work. Section~\ref{sec:conclusion} concludes.

%=====================================================================
\section{Background}
\label{sec:background}
%=====================================================================

\subsection{Zero Trust Architecture Fundamentals}

The NIST Special Publication 800-207~\cite{nist800207} defines the reference architecture for ZTA, comprising three core components: the Policy Engine (PE), which makes trust-based access decisions; the Policy Administrator (PA), which establishes and terminates communication paths; and the Policy Enforcement Point (PEP), which enables and monitors connections between subjects and resources. Trust decisions are informed by multiple signal categories: identity (who is the subject?), device posture (is the device compliant and uncompromised?), network context (where is the request originating?), behavioural analytics (is this consistent with expected patterns?), data sensitivity (what is being accessed?), and threat intelligence (are there known indicators of compromise?).

A critical but often unstated assumption in NIST 800-207 is that the Policy Engine has \textit{continuous visibility} into trust signals and that the network environment is \textit{relatively stable}---that is, the device maintains a persistent connection to the trust infrastructure, and the underlying network characteristics do not change dramatically during a session. Multi-RAT IoT environments violate both assumptions. The PE may lose connectivity to the device during RAT transitions, and the fundamental security properties of the network---authentication model, encryption strength, identity infrastructure---change at each transition.

\subsection{Distinction from 3GPP Handover Security}
\label{sec:3gpp-distinction}

The 3GPP handover security literature is mature and provides well-understood key derivation procedures for intra-family transitions (e.g., 5G to 4G, inter-eNB handovers)~\cite{3gpphandover}. It is important to distinguish these from the trust boundary crossings addressed in this paper. A 3GPP handover operates \textit{within a single trust framework}: the operator's core network manages key continuity, the subscriber identity persists across cells, and the authentication context is preserved via a standardised key hierarchy ($K_{\mathrm{ASME}} \rightarrow K_{\mathrm{eNB}} \rightarrow K_{\mathrm{gNB}}$). By contrast, a multi-RAT trust boundary crossing traverses \textit{between} trust frameworks: the identity model, authentication protocol, key management system, and trust anchor all change simultaneously. There is no overarching core network managing trust continuity between a 5G cell and a LoRaWAN gateway, or between a Meshtastic mesh and an OcuSync link. The 3GPP model demonstrates that intra-framework trust transfer is a solved problem; our contribution addresses the unsolved inter-framework case.

\subsection{Multi-RAT IoT Communication Landscape}

Mobile IoT devices increasingly incorporate multiple radio interfaces to optimise coverage, cost, and resilience. We characterise various RAT families relevant to UAV operations, analysed along trust-relevant dimensions.

\subsubsection{3GPP Cellular (5G/4G LTE)}

Fifth-generation and fourth-generation cellular networks provide the most mature trust infrastructure of any wireless IoT technology. The 5G Authentication and Key Agreement (5G-AKA) protocol provides mutual authentication between the device (via USIM/eSIM) and the network, with subscriber identity privacy via the Subscription Concealed Identifier (SUCI) mechanism. Handover between 5G and 4G within the 3GPP family preserves authentication context through standardised key derivation procedures. However, 5G coverage remains incomplete in rural and remote environments---precisely where many UAV operations occur.

\subsubsection{LoRaWAN}

LoRaWAN is a widely deployed LPWAN protocol operating in unlicensed ISM bands (868\,MHz in Europe, 915\,MHz in North America) with range capabilities of 2--15\,km. LoRaWAN supports two activation modes: Over-the-Air Activation (OTAA), which provides mutual authentication and session key derivation via AES-128, and Activation by Personalisation (ABP), which pre-provisions static keys with no join procedure. LoRaWAN has no concept of device attestation, no handover mechanism between gateways or network servers, and extremely limited bandwidth (0.3--50\,kbps), making in-band trust evidence exchange costly.

\subsubsection{Meshtastic (LoRa Mesh)}

Meshtastic is an open-source mesh networking protocol built atop LoRa physical-layer modulation. Unlike LoRaWAN's star topology with centralised network servers, Meshtastic creates peer-to-peer mesh networks with multi-hop routing. Security relies on AES-256-CTR encryption with a channel-wide pre-shared key (PSK). There is no per-device identity infrastructure, no mutual authentication, no public key infrastructure, and no mechanism for trust evidence exchange. A device joining a Meshtastic mesh proves only knowledge of the channel PSK---it cannot prove its identity, attest its firmware, or present trust evidence from another domain.

\subsubsection{DJI OcuSync (Proprietary)}

DJI's OcuSync protocol (versions 1.0 through O4) is a proprietary transmission system operating in the 2.4\,GHz and 5.8\,GHz ISM bands, providing simultaneous video downlink, telemetry, and command-and-control~\cite{djisecwp,schiller2023dronesec}. OcuSync employs AES-256 encryption with session keys generated via a random number generator at each power-on cycle. Recent enterprise models support mutual authentication using device certificates and private keys stored in ARM TrustZone. DJI also supports 4G-enhanced transmission for extended-range operations, where the link initiates on OcuSync and transitions to cellular---a proprietary multi-RAT handover with no documented trust-state transfer mechanism. Critically, OcuSync is a closed protocol: its authentication procedures, key negotiation, and trust model are not publicly documented, making external trust evaluation and cross-domain trust portability architecturally impossible without vendor cooperation.

Research by Schiller et al.\ demonstrated that DJI's DroneID broadcast---transmitted via both enhanced Wi-Fi and OcuSync---is not encrypted despite vendor claims, revealing drone and operator positions to any receiver with appropriate SDR equipment~\cite{schiller2023dronesec}. This finding underscores the opacity problem: closed protocols cannot be independently verified for trust properties.

\subsubsection{MAVLink (Open Autopilot Telemetry)}

The Micro Air Vehicle Link (MAVLink) protocol is a lightweight, header-only communication protocol used by the two dominant open-source autopilot platforms: ArduPilot and PX4~\cite{mavsec2019}. MAVLink defines bidirectional messages carrying vehicle state, sensor data, and control commands between the UAV and ground control station (GCS). In its most common deployment, MAVLink is transmitted over serial radio links (SiK-firmware radios at 433/915\,MHz, or higher-power RFD900 modems) at data rates of 19.2--250\,kbps.

MAVLink version 1.0 provides \textit{no authentication or encryption whatsoever}---messages are transmitted in cleartext, and any entity capable of transmitting on the same frequency can inject commands~\cite{mavlinkcve2020}. MAVLink 2.0 introduces optional message signing via HMAC-SHA256, providing authentication but not confidentiality, and requires all devices in the network to share a symmetric key. The SiK radio firmware supports 128-bit AES encryption at the link layer, but this is independent of MAVLink and uses a single static key across all radios in a network. From a Zero Trust perspective, MAVLink deployments typically operate with \textit{implicit trust}---the antithesis of ZTA principles.

\subsubsection{Bluetooth Low Energy (BLE)}

BLE is commonly used for close-range communication between UAVs and ground-deployed sensors, payload devices, and configuration interfaces. BLE 4.2+ supports LE Secure Connections with Elliptic Curve Diffie-Hellman (ECDH) key exchange and AES-CCM encryption. However, legacy pairing modes (Just Works, Passkey Entry) remain widely deployed and are vulnerable to eavesdropping and man-in-the-middle attacks~\cite{bleattacks2020}. BLE's trust model is session-based: pairing establishes a bond between two specific devices, but this bond has no relationship to trust established on any other RAT. A drone collecting data from a BLE-connected soil moisture sensor must establish a separate trust relationship with that sensor, independent of its cellular or LoRaWAN trust state. BLE's range limitation (typically 10--50\,m for IoT applications) means that BLE trust boundary crossings are frequent but brief during data collection operations.

\subsubsection{Wi-Fi (IEEE 802.11)}

Wi-Fi provides high-bandwidth, short-range connectivity commonly used for drone operations in close proximity (e.g., DJI consumer models, indoor inspection drones). WPA3/WPA2 provides authentication and encryption, and enterprise Wi-Fi (802.1X with RADIUS) can integrate with identity management systems. However, Wi-Fi coverage is limited and not designed for mobile nodes traversing large areas. Some drone platforms (notably earlier DJI models including the Spark and Mini SE) use Wi-Fi as their primary control link, in some cases with legacy encryption---Christof's reverse engineering revealed WEP encryption on certain DJI Wi-Fi implementations~\cite{christof2021djiwifi}.

\subsubsection{Satellite (LEO/GEO)}

Low Earth Orbit (LEO) satellite connectivity (e.g., Iridium, Starlink, 3GPP NTN) is increasingly available for UAV operations in remote areas. Trust properties vary significantly by provider: Iridium provides proprietary encryption and authentication, Starlink uses WPA2/WPA3 for the user terminal link with proprietary space segment security, and 3GPP Non-Terrestrial Networks (NTN) in Release 17 extend cellular authentication to satellite bearers. Satellite links introduce high latency (20--600\,ms depending on orbit), which affects the feasibility of interactive authentication protocols.

\medskip

Table~\ref{tab:rat-trust} summarises the trust-relevant properties of these seven RAT families.

\begin{table*}[t]
\centering
\caption{Trust-Relevant Properties of IoT/UAV RAT Families}
\label{tab:rat-trust}
\small
\setlength{\tabcolsep}{3pt}
\begin{tabularx}{\textwidth}{@{}l>{\raggedright\arraybackslash}p{1.8cm}>{\raggedright\arraybackslash}p{1.6cm}>{\raggedright\arraybackslash}p{1.6cm}>{\raggedright\arraybackslash}p{1.6cm}>{\raggedright\arraybackslash}p{1.6cm}>{\raggedright\arraybackslash}p{1.6cm}>{\raggedright\arraybackslash}p{1.7cm}>{\raggedright\arraybackslash}p{2.5cm}@{}}
\toprule
\textbf{Property} & \textbf{5G/4G} & \textbf{LoRa\-WAN} & \textbf{Mesh\-tastic} & \textbf{OcuSync} & \textbf{MAV\-Link} & \textbf{BLE} & \textbf{Wi-Fi} & \textbf{Satellite} \\
\midrule
Auth. & 5G-AKA (mutual) & OTAA/ ABP & PSK & Propr.\ (cert+TZ) & None/ HMAC & LE SC/ legacy & WPA2/3 & Iridium: propr.; NTN: 5G-AKA \\
\addlinespace
Identity & SUPI/ IMSI & DevEUI & Node ID & Serial (closed) & Sys.\ ID & MAC/ IRK & MAC/ 802.1X & Iridium: SIM; Starlink: term.\ cert \\
\addlinespace
Encrypt. & 128/256 & AES-128 & AES-256 & AES-256 & None (opt.) & AES-CCM & AES-CCMP & Iridium: AES-256; NTN: 5G enc. \\
\addlinespace
Mutual & Yes & OTAA & No & Yes (ent.) & No & LE SC & Ent.\ only & NTN: yes; others: varies \\
\addlinespace
Handover & 3GPP & None & None & Propr.\ (4G) & None & None & Roaming & NTN: 3GPP; LEO: limited \\
\addlinespace
Anchor & USIM & Net.\ Srv. & PSK & TrustZone & HMAC key & Bond & AP/RAD. & Iridium: SIM; Starlink: cert \\
\addlinespace
BW & High & V.\ low & V.\ low & High & Low & Med. & High & Low--Med \\
\addlinespace
Open & Std. & Open std. & Open src. & Closed & Open src. & Std. & Std. & NTN: std.; others: closed \\
\bottomrule
\end{tabularx}
\setlength{\tabcolsep}{6pt}
\end{table*}

\subsection{UAVs as Multi-RAT Platforms}

Commercial and enterprise UAVs increasingly operate as multi-RAT platforms. A modern DJI Matrice series drone, for example, simultaneously operates an OcuSync link for command and video, a 4G cellular dongle for extended-range operations, and may interact with LoRaWAN ground sensors as part of a data collection mission. Open-source platforms built on ArduPilot or PX4 commonly carry a SiK or RFD900 radio for MAVLink telemetry alongside cellular modems and LoRa modules. In defence and emergency services contexts, additional military-grade radios and mesh networking capabilities further expand the RAT portfolio.

This multi-RAT capability makes UAVs uniquely challenging for ZTA implementation. Unlike a stationary IoT sensor that connects to a single LPWAN gateway for its operational lifetime, a drone traverses multiple coverage areas, transitions between RATs dynamically based on availability and mission requirements, and must maintain security posture throughout. The drone is simultaneously power-constrained (every milliwatt consumed by security reduces flight time), mission-critical (disruption has operational consequences), and high-value (making it an attractive target for adversarial manipulation)~\cite{ieeespectrumzt2023}.

\subsection{Regulatory Context: Remote Identification}
\label{sec:remoteid}

The regulatory environment adds a further trust-relevant signal. The European Union Delegated Regulation 2019/945~\cite{eudroneregulation} and the United States FAA Remote ID Rule (14 CFR Part 89, effective March 2024)~\cite{faaremoteid} mandate that UAVs broadcast identification and location information during flight. Remote ID operates as a continuous identity assertion: the drone broadcasts its serial number (or session ID), position, altitude, velocity, and operator location.

From a ZTA perspective, Remote ID presents a paradox. It provides a persistent identity signal---broadcast continuously regardless of RAT transitions---yet this signal is \textit{unauthenticated} in most implementations. Standard Remote ID (broadcast via Wi-Fi Neighbour Awareness Networking or Bluetooth 4.0/5.0 advertising) transmits in cleartext with no cryptographic binding to the device's actual identity. An adversary can trivially spoof Remote ID broadcasts. The ASTM F3411-22a standard~\cite{astmf3411} does not mandate message authentication or integrity protection.

This creates a trust signal that is universally available but cryptographically worthless under ZTA principles. Remote ID can serve as a \textit{weak contextual signal}---if the device's self-reported Remote ID position matches its position as derived from authenticated sources, this provides marginal corroborative evidence---but it cannot substitute for proper authentication. We incorporate Remote ID into the contextual trust component of our model with appropriately low weight.

%=====================================================================
\section{Related Work}
\label{sec:related}
%=====================================================================

\subsection{Zero Trust for IoT and UAVs}

The application of ZTA principles to IoT and UAV systems has received growing attention. Alquwayzani and Albuali~\cite{alquwayzani2024slr} present a systematic literature review of ZTA for UAV security in the Internet of Battlefield Things (IoBT), surveying authentication, authorisation, and access control mechanisms. Their review identifies continuous verification and microsegmentation as key ZTA components for UAV security, but does not address the trust implications of multi-RAT network transitions.

A blockchain-based continuous authentication approach for UAVs operating in 5G environments was proposed in~\cite{springer2025blockchain5guav}, using blockchain's decentralisation and transparency to maintain real-time verification. However, the work assumes a single-RAT (5G) environment and does not consider scenarios where the UAV transitions to non-cellular networks.

Saghaian et al.~\cite{saghaian2023bayesian} develop a risk-based Zero Trust approach for tactical edge computing using a Bayesian Network model for communication request risk evaluation. Their framework adapts verification rigour to resource constraints and scenario risk levels, representing the closest existing work to our approach. However, their model treats the network environment as a parameter within a single trust domain rather than modelling transitions \textit{between} trust domains.

Ramezanpour and Jagannath~\cite{ramezanpour2022izta} propose an intelligent ZTA framework for 5G/6G networks using machine learning within the Open Radio Access Network (O-RAN) architecture. While they address the heterogeneity of 5G/6G services, the analysis remains within the 3GPP family and does not extend to non-cellular RATs.

The lightweight FAST-SM9 Zero Trust authentication framework~\cite{fastsm92025} addresses distributed authentication across Cloud-Edge-End layers in IoT, achieving low-latency authentication in heterogeneous environments. While the work addresses network heterogeneity, it does not model RAT-level trust state transitions or the power cost of re-authentication.

\subsection{Multi-RAT Handover and IoT Connectivity}

The multi-RAT IoT connectivity literature focuses primarily on performance optimisation rather than security. Leenders et al.~\cite{leenders2021multirat} evaluate LoRaWAN and NB-IoT in a multi-RAT configuration, demonstrating energy efficiency improvements of at least 4$\times$ for use cases with sporadic large payloads. Their analysis is performance-focused and does not examine trust implications of RAT switching.

The most directly relevant existing work is ``Enabling Roaming Across Heterogeneous IoT Wireless Networks: LoRaWAN MEETS 5G''~\cite{lorawan5groaming2020}, which proposes a handover roaming mechanism using the 5G network as a trust anchor for LoRaWAN device authentication and key management. This paper takes a significant step towards cross-RAT trust transfer but does not frame the problem in Zero Trust terms, does not generalise beyond two RATs, and does not consider the power cost of trust re-establishment.

Sanchez-Iborra et al.~\cite{sancheziborra2019multirat} demonstrate multi-RAT 5G network slicing for the Internet of Vehicles, enabling simultaneous LoRaWAN and cellular connectivity with QoS-differentiated slices. The work addresses performance isolation but does not analyse trust boundaries between slices.

\subsection{Drone Communication Security}

The security of drone communication protocols has been extensively studied. Allouch et al.~\cite{mavsec2019} identify MAVLink's security vulnerabilities and propose MAVSec, integrating encryption algorithms (AES-CBC, AES-CTR, RC4, ChaCha20) into the ArduPilot implementation. The MAVLink protocol's complete lack of authentication in version 1.0 was formally catalogued as CVE-2020-10282~\cite{mavlinkcve2020}, with a CVSS score of 9.8 (critical).

Schiller et al.~\cite{schiller2023dronesec} present a comprehensive security analysis of DJI drones, reverse-engineering the OcuSync protocol and demonstrating that DroneID broadcasts are unencrypted despite vendor claims. Their work also develops a black-box fuzzer discovering security-critical crashes in DJI firmware. These findings underscore that proprietary drone protocols, despite vendor security claims, may harbour undisclosed vulnerabilities that undermine trust assumptions.

\subsection{Research Gap}

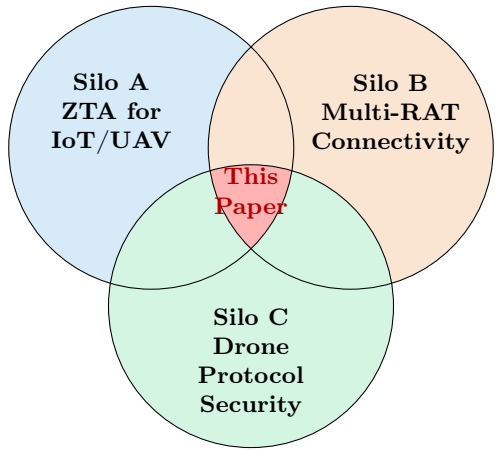
\begin{figure*}[t]
\centering
\begin{tikzpicture}[scale=0.75]
    % Circles
    \fill[siloA!20] (0,0) circle (2.5cm);
    \fill[siloB!20] (3.5,0) circle (2.5cm);
    \fill[siloC!20] (1.75,-2.8) circle (2.5cm);
    % Intersection highlight
    \begin{scope}
        \clip (0,0) circle (2.5cm);
        \clip (3.5,0) circle (2.5cm);
        \clip (1.75,-2.8) circle (2.5cm);
        \fill[red!30] (-1,-4) rectangle (5,3);
    \end{scope}
    % Labels
    \draw (0,0) circle (2.5cm);
    \draw (3.5,0) circle (2.5cm);
    \draw (1.75,-2.8) circle (2.5cm);
    \node[font=\small\bfseries,text width=2.2cm,align=center] at (-0.7,0.6) {Silo A\\ZTA for\\IoT/UAV};
    \node[font=\small\bfseries,text width=2.4cm,align=center] at (4.2,0.6) {Silo B\\Multi-RAT\\Connectivity};
    \node[font=\small\bfseries,text width=2.4cm,align=center] at (1.75,-3.8) {Silo C\\Drone Protocol\\Security};
    \node[font=\small\bfseries,text=red!70!black,text width=2cm,align=center] at (1.75,-0.8) {\textbf{This\\Paper}};
\end{tikzpicture}
\caption{Research gap at the intersection of three established research silos. No prior work addresses Zero Trust trust-state management across heterogeneous RAT transitions including proprietary drone protocols.}
\label{fig:gap}
\end{figure*}

The existing literature addresses Zero Trust for IoT/UAVs, multi-RAT connectivity optimisation, and drone protocol security in separate silos (Figure~\ref{fig:gap}). No prior work systematically examines what happens to Zero Trust compliance when a device transitions between fundamentally different radio access technologies. Specifically:

\begin{itemize}[nosep]
    \item \textbf{Silo A} (ZTA for IoT/UAV) assumes a single or stable network environment.
    \item \textbf{Silo B} (Multi-RAT IoT connectivity) addresses performance but ignores trust-state security.
    \item \textbf{Silo C} (Drone protocol security) analyses individual protocols but not cross-protocol trust transitions.
\end{itemize}

Our contribution occupies the intersection: ZTA trust-state management across heterogeneous RAT transitions in resource-constrained mobile IoT, with explicit treatment of proprietary drone protocols as opaque trust domains.

%=====================================================================
\section{Trust Boundary Model for Multi-RAT IoT}
\label{sec:model}
%=====================================================================

This section presents the core conceptual contribution: a formal model of trust state and trust boundary crossings in multi-RAT environments.

\subsection{Trust State}

\begin{definition}[Trust State]
\label{def:truststate}
The \textit{trust state} of device $d$ at time $t$ operating on RAT $R$ is a five-component vector:
\[
\mathbf{s}(d, t, R) = \langle s_{\mathrm{id}}, s_{\mathrm{dev}}, s_{\mathrm{ctx}}, s_{\mathrm{net}}, s_{\mathrm{pol}} \rangle \in [0,1]^5
\]
where the components represent:
\begin{itemize}[nosep]
    \item $s_{\mathrm{id}}$: \textbf{Identity trust} --- strength of device authentication $\times$ verification recency.
    \item $s_{\mathrm{dev}}$: \textbf{Device trust} --- firmware attestation level $\times$ posture compliance.
    \item $s_{\mathrm{ctx}}$: \textbf{Contextual trust} --- behavioural confidence $\times$ environmental consistency.
    \item $s_{\mathrm{net}}$: \textbf{Network trust} --- channel encryption strength $\times$ infrastructure integrity.
    \item $s_{\mathrm{pol}}$: \textbf{Policy trust} --- enforcement completeness $\times$ audit coverage.
\end{itemize}
\end{definition}

Each component captures a distinct dimension of the Zero Trust verification model. Identity trust reflects authentication; device trust reflects attestation; contextual trust reflects behavioural and environmental signals; network trust reflects the security properties of the communication channel itself; and policy trust reflects the degree to which ZTA policies can be enforced and audited on the current RAT.

\begin{definition}[Composite Trust Score]
\label{def:composite}
The composite trust score is a weighted aggregation:
\[
\mathcal{T}(d, t, R) = \sum_{j \in \mathcal{J}} w_j \cdot s_j(d, t, R)
\]
where $\mathcal{J} = \{\mathrm{id}, \mathrm{dev}, \mathrm{ctx}, \mathrm{net}, \mathrm{pol}\}$, weights $w_j > 0$ satisfy $\sum_{j} w_j = 1$, and the weight vector $\mathbf{w}$ is determined by organisational risk policy $\Pi$.
\end{definition}

The weight vector $\mathbf{w}$ encodes the organisation's security priorities. A defence application might weight identity and device trust heavily (prioritising authentication and attestation in contested environments), while a commercial IoT deployment might prioritise network and contextual trust (prioritising data integrity and operational normality).

\subsection{Weight Vector Selection}
\label{sec:weights}

The weight vector $\mathbf{w}$ encodes the organisation's security priorities and is a policy decision, not a technical one. However, the model imposes a structural requirement: weights should reflect the \textit{relative importance} of each trust component to the operational risk profile. We illustrate with two representative profiles:

\textbf{Defence / contested environment}: $\mathbf{w}_{\mathrm{def}} = (0.30, 0.25, 0.15, 0.15, 0.15)$. Identity and device trust are weighted heavily because adversary impersonation and compromised devices are primary threats. Network trust receives lower weight because the network is assumed hostile by doctrine.

\textbf{Commercial IoT}: $\mathbf{w}_{\mathrm{com}} = (0.20, 0.15, 0.20, 0.25, 0.20)$. Network trust receives higher weight because commercial operations rely on network integrity for data assurance. Contextual trust is weighted higher because anomaly detection is a primary indicator of compromise.

A further design question is whether $\mathbf{w}$ should vary with the active RAT. The argument for RAT-dependent weights is that the achievable trust profile differs by RAT: weighting network trust at 0.25 on Meshtastic (maximum achievable $s_{\mathrm{net}} \approx 0.3$) means network trust can never contribute more than 0.075 to the composite score. The counter-argument is that RAT-dependent weights complicate policy specification and introduce potential gaming (an adversary who can influence RAT selection can influence which weight vector applies). We recommend static weight vectors with RAT-dependent \textit{trust ceilings} (Section~\ref{sec:discussion}) as the cleaner architectural choice.

\subsection{Temporal Trust Decay}
\label{sec:decay}

Between transitions, trust does not remain static. An authentication performed 30 minutes ago provides less assurance than one performed 30 seconds ago, because the device's state may have changed without the PE's knowledge.

\begin{definition}[Trust Decay]
\label{def:decay}
For trust component $j$, the value decays over time on a stable RAT $R$ according to:
\[
s_j(d, t_0 + \Delta t, R) = s_j(d, t_0, R) \cdot e^{-\lambda_j(R) \cdot \Delta t}
\]
where $\lambda_j(R) \geq 0$ is the decay rate for component $j$ on RAT $R$, and $t_0$ is the time of last verification.
\end{definition}

The exponential decay model is a first-order approximation chosen for analytical tractability. Alternative decay profiles may better capture specific scenarios: the Weibull distribution $s_j(t_0 + \Delta t) = s_j(t_0) \cdot e^{-(\lambda_j \Delta t)^{k_j}}$ with shape parameter $k_j$ enables modelling of non-exponential dynamics---$k_j < 1$ produces rapid initial decay that slows over time (appropriate for volatile trust signals such as behavioural context), while $k_j > 1$ produces slow initial decay that accelerates (appropriate for trust backed by time-limited credentials approaching expiry). Step-function decay triggered by specific events (e.g., certificate revocation, firmware update notification) may also be appropriate for certain components. The exponential model ($k_j = 1$) captures the general intuition that trust degrades continuously but is refreshed by re-verification events. The decay rate $\lambda_j(R)$ is RAT-dependent: a 5G network with continuous monitoring infrastructure supports slower decay (lower $\lambda$) than a Meshtastic mesh with no monitoring capability (higher $\lambda$). Identity trust decays slowly when the authentication protocol provides session continuity (e.g., 5G with active security context) and rapidly when there is no session mechanism (e.g., MAVLink). Device trust decays slowly for hardware-attested devices (TPM quotes have long validity) and rapidly for software-only posture checks.

The interaction between trust decay and boundary crossings produces the sawtooth pattern observed in multi-RAT missions: trust decays gradually between transitions and drops sharply at each transition, with partial recovery after re-authentication on the new RAT.

\subsection{Taxonomy of RAT Transitions}
\label{sec:taxonomy}

Not all trust boundary crossings are equivalent. We identify four categories with distinct trust implications:

\begin{definition}[Transition Taxonomy]
\label{def:taxonomy}
RAT transitions are classified as:
\begin{itemize}[nosep]
    \item \textbf{Planned ($\beta^P$)}: Mission-scripted transitions at known locations and times. The device can pre-stage trust evidence on the target RAT. Trust recovery is fastest.
    \item \textbf{Coverage-driven ($\beta^C$)}: Signal loss on the current RAT forces fallback. Predictable in character but unpredictable in timing. Limited pre-staging may be possible if signal degradation is detectable.
    \item \textbf{Opportunistic ($\beta^O$)}: A higher-capability or lower-cost RAT becomes available. The device maintains the existing RAT during transition, enabling parallel authentication before committing.
    \item \textbf{Adversary-forced ($\beta^A$)}: Jamming, spoofing, or denial-of-service compels an unplanned switch. No preparation time, the fallback RAT may be adversary-selected, and the trust state of the previous domain may be compromised.
\end{itemize}
\end{definition}

Table~\ref{tab:transition-types} summarises the properties of each transition type. The cost multiplier $\alpha(\beta)$ captures the overhead: planned transitions ($\alpha = 1.0$) incur baseline cost, while adversary-forced transitions ($\alpha = 2.0$) incur doubled cost due to adversary-resilient re-authentication requirements.

\begin{table*}[t]
\centering
\caption{Properties of RAT Transition Types}
\label{tab:transition-types}
\small
\begin{tabular}{@{}lcccc@{}}
\toprule
\textbf{Property} & \textbf{Planned ($\beta^P$)} & \textbf{Coverage ($\beta^C$)} & \textbf{Opport.\ ($\beta^O$)} & \textbf{Adversary ($\beta^A$)} \\
\midrule
Preparation time & High & Low--Med & Medium & None \\
Target RAT known & Yes & Partially & Yes & No \\
Pre-staging possible & Yes & Limited & Yes (parallel) & No \\
Trust gap duration & Minimal & Short & Minimal & Extended \\
Adversarial risk & Low & Low & Medium & High \\
Cost multiplier $\alpha$ & $1.0\times$ & $1.3\times$ & $0.8\times$ & $2.0\times$ \\
\bottomrule
\end{tabular}
\end{table*}

\subsection{Trust Boundary Crossing Events}

\begin{definition}[Trust Boundary Crossing]
\label{def:boundary}
A \textit{trust boundary crossing event} $\beta_{i \rightarrow i+1}$ occurs when device $d$ transitions from RAT $R_i$ to RAT $R_{i+1}$ at time $t_\beta$. The event induces a trust state transformation:
\[
\mathbf{s}(d, t_\beta^{-}, R_i) \xrightarrow{\;\beta_{i \rightarrow i+1}\;} \mathbf{s}(d, t_\beta^{+}, R_{i+1})
\]
where $t_\beta^{-}$ denotes the instant before transition and $t_\beta^{+}$ denotes the instant after.
\end{definition}

The key question is: what fraction of each trust component survives the boundary crossing?

\begin{definition}[Trust Survival Function]
\label{def:survival}
For trust component $j \in \mathcal{J}$, the \textit{trust survival function} $\sigma_j : \mathcal{R} \times \mathcal{R} \rightarrow [0,1]$ defines the proportion of trust that persists across a RAT transition:
\[
s_j(d, t_\beta^{+}, R_{i+1}) = \sigma_j(R_i, R_{i+1}) \cdot s_j(d, t_\beta^{-}, R_i)
\]
where $\mathcal{R}$ is the set of all RATs.
\end{definition}

When $\sigma_j(R_i, R_{i+1}) = 0$, trust component $j$ is completely invalidated by the transition---full re-establishment is required. When $\sigma_j(R_i, R_{i+1}) = 1$, the trust component is fully preserved. Values in $(0,1)$ indicate partial trust survival, where some evidence carries over but confidence is reduced.

\subsection{Trust Survival Analysis Across RAT Families}

We now qualitatively characterise the trust survival functions for each component across the seven RAT families. These values represent reasoned assessments based on protocol analysis; empirical measurement on specific hardware platforms is identified as future work.

\subsubsection{Identity Trust Survival ($\sigma_{\mathrm{id}}$)}

Identity trust depends on whether the authentication credential used on $R_i$ can be verified or recognised by $R_{i+1}$'s trust infrastructure. Table~\ref{tab:sigma-id} presents the identity trust survival matrix.

\begin{table*}[t]
\centering
\caption{Identity Trust Survival Matrix ($\sigma_{\mathrm{id}}$). Values represent the fraction of identity trust preserved when transitioning from the row RAT to the column RAT. A value of 0.0 indicates complete identity trust loss requiring full re-authentication.}
\label{tab:sigma-id}
\small
\begin{tabular}{@{}lccccccc@{}}
\toprule
\textbf{From $\backslash$ To} & \textbf{5G} & \textbf{LoRa\-WAN} & \textbf{Mesh} & \textbf{OcuSync} & \textbf{MAV\-Link} & \textbf{Wi-Fi} & \textbf{Sat.} \\
\midrule
5G/4G     & 0.95 & 0.0  & 0.0  & 0.0  & 0.0  & 0.3  & 0.3  \\
LoRaWAN   & 0.0  & 0.9  & 0.0  & 0.0  & 0.0  & 0.0  & 0.0  \\
Meshtastic & 0.0  & 0.0  & 0.7  & 0.0  & 0.0  & 0.0  & 0.0  \\
OcuSync   & 0.0  & 0.0  & 0.0  & 0.85 & 0.0  & 0.0  & 0.0  \\
MAVLink   & 0.0  & 0.0  & 0.0  & 0.0  & 0.8  & 0.0  & 0.0  \\
Wi-Fi     & 0.3  & 0.0  & 0.0  & 0.0  & 0.0  & 0.9  & 0.0  \\
Satellite & 0.3  & 0.0  & 0.0  & 0.0  & 0.0  & 0.0  & 0.85 \\
\bottomrule
\end{tabular}
\end{table*}

Several observations follow from this matrix. First, cross-family identity trust survival is overwhelmingly zero. A 5G-AKA authentication bears no relationship to a LoRaWAN OTAA join, an OcuSync session key, or a MAVLink HMAC credential. Second, within-family survival is high but not unity, reflecting the cost of session re-keying, handover procedures, and re-association. Third, the only non-zero cross-family values occur between 5G and Wi-Fi (where 802.1X enterprise authentication can share an identity infrastructure with cellular via the Evolved Packet Core) and between 5G and satellite (where 3GPP NTN extends the cellular trust model to non-terrestrial bearers). Fourth, proprietary protocols (OcuSync) and weakly authenticated protocols (MAVLink, Meshtastic) form isolated trust islands---no external trust evidence can enter or leave.

\subsubsection{Device Trust Survival ($\sigma_{\mathrm{dev}}$)}

Device trust---firmware attestation, hardware integrity, posture compliance---has the highest portability across RATs because it is fundamentally independent of the communication channel. A TPM attestation result is a property of the device, not the network. Table~\ref{tab:sigma-dev} reflects this.

\begin{table*}[t]
\centering
\caption{Device Trust Survival Matrix ($\sigma_{\mathrm{dev}}$). Device attestation evidence is RAT-independent when based on hardware roots of trust, yielding uniformly high survival values.}
\label{tab:sigma-dev}
\small
\begin{tabular}{@{}lccccccc@{}}
\toprule
\textbf{From $\backslash$ To} & \textbf{5G} & \textbf{LoRa\-WAN} & \textbf{Mesh} & \textbf{OcuSync} & \textbf{MAV\-Link} & \textbf{Wi-Fi} & \textbf{Sat.} \\
\midrule
5G/4G     & 0.95 & 0.7  & 0.6  & 0.5  & 0.5  & 0.8  & 0.8  \\
LoRaWAN   & 0.7  & 0.95 & 0.6  & 0.5  & 0.5  & 0.7  & 0.7  \\
Meshtastic & 0.5  & 0.5  & 0.85 & 0.4  & 0.4  & 0.5  & 0.5  \\
OcuSync   & 0.5  & 0.5  & 0.4  & 0.90 & 0.4  & 0.5  & 0.5  \\
MAVLink   & 0.5  & 0.5  & 0.4  & 0.4  & 0.85 & 0.5  & 0.5  \\
Wi-Fi     & 0.8  & 0.7  & 0.5  & 0.5  & 0.5  & 0.95 & 0.7  \\
Satellite & 0.8  & 0.7  & 0.5  & 0.5  & 0.5  & 0.7  & 0.95 \\
\bottomrule
\end{tabular}
\end{table*}

Device trust survival is substantially higher than identity trust survival across all RAT pairs. The variation reflects not the attestation evidence itself (which is RAT-independent) but the \textit{receiving domain's ability to verify} that evidence. A 5G network with an integrated RATS (Remote Attestation Procedures) verifier can consume an Entity Attestation Token (EAT)~\cite{rfc9334}; a Meshtastic mesh cannot. The reduced values for proprietary protocols (OcuSync) reflect the closed ecosystem's inability to accept external attestation results.

\subsubsection{Network Trust Survival ($\sigma_{\mathrm{net}}$)}

Network trust is fundamentally non-portable. Channel encryption, network integrity guarantees, and infrastructure trust are properties of the specific RAT. When a device transitions from 5G to LoRaWAN, the 5G channel's encryption provides no protection on the LoRaWAN link. Accordingly, $\sigma_{\mathrm{net}}(R_i, R_{i+1}) \approx 0$ for all cross-family transitions, with modest values only for intra-family handovers (e.g., 5G $\rightarrow$ 4G: $\sigma_{\mathrm{net}} \approx 0.8$). This represents the irreducible cost of multi-RAT ZTA compliance.

\subsubsection{Contextual and Policy Trust Survival}

Contextual trust ($\sigma_{\mathrm{ctx}}$) exhibits mixed portability. Device-local context (GPS position, IMU data, flight telemetry) is fully portable---the device generates this data independently of the RAT. Network-derived context (traffic patterns, peer behaviour, signal environment) is not portable. Policy trust ($\sigma_{\mathrm{pol}}$) depends on whether the new RAT's infrastructure can enforce and audit the same policy set. Highly capable RATs (5G, enterprise Wi-Fi) can enforce rich policies; constrained RATs (LoRaWAN, Meshtastic, MAVLink) support only minimal policy enforcement.

\subsection{Trust Recovery Cost}

\begin{definition}[Trust Recovery Cost]
\label{def:recoverycost}
The trust recovery cost at boundary crossing $\beta_{i \rightarrow i+1}$ is:
\[
C(\beta_{i \rightarrow i+1}) = \alpha(\beta) \cdot \sum_{j \in \mathcal{J}} c_j(R_{i+1}) \cdot \bigl(1 - \sigma_j(R_i, R_{i+1})\bigr)
\]
where $c_j(R)$ denotes the cost (in joules) of fully re-establishing trust component $j$ on RAT $R$, and $\alpha(\beta)$ is the transition type cost multiplier from Table~\ref{tab:transition-types}.
\end{definition}

The term $(1 - \sigma_j)$ captures the \textit{trust deficit}---the fraction of trust that must be rebuilt. When survival is high, recovery cost is low; when survival is zero, the full cost $c_j(R)$ is incurred. The cumulative trust recovery cost over a mission with $n$ RAT transitions is:
\begin{equation}
C_{\mathrm{mission}} = \sum_{k=1}^{n-1} C(\beta_{k \rightarrow k+1})
\label{eq:missioncost}
\end{equation}

For a device with total power budget $P_{\max}$ and mission duration $\Delta t$, the power available for authentication is constrained by:
\begin{equation}
P_{\mathrm{auth}} \leq P_{\max} - P_{\mathrm{flight}} - P_{\mathrm{payload}} - P_{\mathrm{comms}}
\label{eq:powerbudget}
\end{equation}

If $C_{\mathrm{mission}} > P_{\mathrm{auth}} \cdot \Delta t$, the device \textit{cannot sustain full Zero Trust compliance across all RAT transitions}. This fundamental tension between ZTA mandates and resource constraints is the central finding of this paper and the motivation for power-aware trust optimisation.

\subsection{Worked Numerical Example}
\label{sec:worked-example}

We illustrate the complete model with a single coverage-driven transition ($\alpha = 1.3$): a drone on 4G LTE transitioning to LoRaWAN.

\begin{tcolorbox}[colback=examplebox,colframe=trusthigh,title=\textbf{Worked Example: 4G $\rightarrow$ LoRaWAN Trust Boundary Crossing},breakable]
\textbf{Step 1: Pre-transition trust state on 4G.}

Using commercial weight vector $\mathbf{w}_{\mathrm{com}} = (0.20, 0.15, 0.20, 0.25, 0.20)$:
\[
\mathbf{s}(d, t_\beta^{-}, \mathrm{4G}) = \langle 0.88, 0.82, 0.75, 0.85, 0.78 \rangle
\]
\begin{multline*}
\mathcal{T}_{\mathrm{pre}} = 0.20(0.88) + 0.15(0.82) + 0.20(0.75) \\
+ 0.25(0.85) + 0.20(0.78) = 0.821
\end{multline*}

\textbf{Step 2: Apply trust survival functions.}

From the survival matrices (identity from Table~\ref{tab:sigma-id}, device from Table~\ref{tab:sigma-dev}, network $\sigma_{\mathrm{net}} = 0.0$, contextual $\sigma_{\mathrm{ctx}} = 0.5$, policy $\sigma_{\mathrm{pol}} = 0.2$):
\[
\boldsymbol{\sigma}(\mathrm{4G}, \mathrm{LoRaWAN}) = \langle 0.0,\; 0.7,\; 0.5,\; 0.0,\; 0.2 \rangle
\]

Post-transition: $s_{\mathrm{id}}' = 0.0$, $s_{\mathrm{dev}}' = 0.7 \times 0.82 = 0.574$, $s_{\mathrm{ctx}}' = 0.5 \times 0.75 = 0.375$, $s_{\mathrm{net}}' = 0.0$, $s_{\mathrm{pol}}' = 0.2 \times 0.78 = 0.156$.
\begin{multline*}
\mathcal{T}_{\mathrm{post}} = 0.20(0.0) + 0.15(0.574) + 0.20(0.375) \\
+ 0.25(0.0) + 0.20(0.156) = 0.242
\end{multline*}

\textbf{Trust drop}: $\Delta\mathcal{T} = 0.821 - 0.242 = 0.579$ (70.5\% reduction).

\medskip
\textbf{Step 3: Trust recovery cost.}

Using per-component re-establishment costs $c_j(\mathrm{LoRaWAN}) = \langle 350, 180, 120, 200, 80 \rangle$\,mJ:
\begin{multline*}
C(\beta) = 1.3 \times [350 + 54 + 60 + 200 + 64] \\
= 1.3 \times 728 = \mathbf{946\,\textbf{mJ}}
\end{multline*}

\textbf{Step 4: Post-recovery trust state.}

After full re-authentication on LoRaWAN (OTAA join), trust is bounded by LoRaWAN's capabilities:
\[
\mathbf{s}_{\mathrm{recovered}} = \langle 0.65, 0.574, 0.55, 0.45, 0.35 \rangle
\]
\begin{multline*}
\mathcal{T}_{\mathrm{recovered}} = 0.20(0.65) + 0.15(0.574) \\
+ 0.20(0.55) + 0.25(0.45) + 0.20(0.35) = 0.499
\end{multline*}

\textbf{Key finding}: Even after full re-authentication at a cost of 946\,mJ, the recovered trust score (0.50) remains \textit{below} $\mathcal{T}_{\min} = 0.6$ because LoRaWAN's trust ceiling is inherently lower than 4G's. The device cannot achieve the minimum trust threshold on this RAT regardless of authentication effort expended.
\end{tcolorbox}

%=====================================================================
\section{Trust Portability}
\label{sec:portability}
%=====================================================================

The trust survival analysis in Section~\ref{sec:model} reveals that most cross-RAT transitions result in near-total trust loss, requiring expensive re-authentication. This section introduces \textit{trust portability}---mechanisms that enable trust evidence to survive RAT transitions, thereby improving survival values and reducing recovery costs.

\subsection{Definition}

\begin{definition}[Trust Portability]
\label{def:portability}
Trust component $s_j$ is \textit{portable} across RAT transition $R_i \rightarrow R_{i+1}$ if there exists a trust evidence artefact $\epsilon_j$ produced on $R_i$ such that:
\begin{enumerate}[nosep]
    \item \textbf{Cost reduction}: $c_{\mathrm{verify}}(\epsilon_j, R_{i+1}) < c_j(R_{i+1})$ --- verifying the artefact on the new RAT is cheaper than full re-establishment.
    \item \textbf{Trust improvement}: $\sigma_j(R_i, R_{i+1} \mid \epsilon_j) > \sigma_j(R_i, R_{i+1})$ --- trust survival is higher when evidence is presented.
    \item \textbf{Freshness}: $\epsilon_j$ includes a timestamp or nonce preventing replay.
    \item \textbf{Integrity}: $\epsilon_j$ is cryptographically bound to the originating trust assessment such that modification is detectable.
\end{enumerate}
\end{definition}

Beyond these functional requirements, portable trust artefacts must satisfy formal security properties:

\begin{definition}[Trust Artefact Security Properties]
\label{def:artefact-security}
A well-formed trust artefact $\epsilon_j$ must satisfy:
\begin{itemize}[nosep]
    \item \textbf{Unforgeability}: No entity other than the original trust assessor can produce a valid $\epsilon_j$. Formally, for any probabilistic polynomial-time adversary $\mathcal{A}$, $\Pr[\mathcal{A} \text{ forges valid } \epsilon_j] \leq \mathrm{negl}(\kappa)$ where $\kappa$ is the security parameter.
    \item \textbf{Non-transferability}: $\epsilon_j$ is cryptographically bound to device $d$ such that it cannot be used by device $d' \neq d$. Achieved by binding the artefact to the device's hardware identity (e.g., TPM-sealed key, DICE identity).
    \item \textbf{Scope limitation}: $\epsilon_j$ does not grant broader access or higher trust than the original assessment warranted.
    \item \textbf{Revocability}: There exists a mechanism to invalidate $\epsilon_j$ before its natural expiry, enabling response to detected compromise. This is challenging in disconnected environments and represents an open problem.
    \item \textbf{Minimal disclosure}: $\epsilon_j$ reveals only the trust assessment result, not the underlying evidence. The receiving domain learns that the device achieved a certain trust level, not the raw attestation data.
\end{itemize}
\end{definition}

Trust portability is not binary---it admits degrees. A hardware-anchored attestation token verified by a well-known root of trust achieves high portability (conditions 1--4 strongly satisfied); a self-signed assertion achieves low portability (conditions 2 and 3 weakly satisfied, condition 4 dependent on trust in the self-signer).

\subsection{Portability Mechanisms by Trust Component}

\subsubsection{Identity Trust Portability}

Four candidate mechanisms exist for carrying identity trust across RAT boundaries:

\textbf{Token-based identity}: A Concise Binary Object Representation (CBOR) Web Token (CWT) or JSON Web Token (JWT) signed by the previous domain's Policy Decision Point. The receiving domain can verify the signature if it trusts the issuer. This requires either a pre-established cross-domain trust relationship or a shared trust anchor (e.g., a common certificate authority). For IoT, CWT is preferred due to its compact encoding.

\textbf{Certificate-based identity}: X.509 certificates or lightweight IoT certificates (e.g., CBOR certificates per the IETF C509 specification). The device presents its certificate on the new RAT, and the receiving infrastructure verifies it against a known root. This works where PKI infrastructure is available on both sides---feasible for 5G-to-Wi-Fi transitions, infeasible for LoRaWAN-to-Meshtastic transitions.

\textbf{Decentralised identifiers (DIDs)}: Self-sovereign identity using W3C Decentralised Identifiers, where the device holds a DID verifiable via a distributed ledger or verifiable data registry. Promising in principle but requires connectivity to the resolution infrastructure, which may not be available during RAT transitions.

\textbf{Hardware-anchored identity}: The device's identity is bound to a hardware root of trust (TPM, DICE, ARM TrustZone), producing attestation evidence that is intrinsically RAT-independent. This is the most promising mechanism for constrained devices, as the trust anchor exists on the device itself rather than in external infrastructure.

\subsubsection{Device Trust Portability}

Device attestation evidence has the highest inherent portability. The IETF Remote Attestation Procedures (RATS) architecture~\cite{rfc9334} defines Entity Attestation Tokens (EATs) that encapsulate device state claims. An EAT produced during 5G-authenticated operation can, in principle, be presented on a LoRaWAN network as evidence of device integrity. The practical challenge is that LoRaWAN's extremely limited bandwidth (a LoRaWAN uplink at SF12 carries only 51 bytes per message) constrains the size of attestation evidence that can be transmitted in-band.

\subsubsection{Contextual Trust Portability}

Device-generated context (GPS, IMU, flight telemetry) is fully portable because it is independent of the RAT. A signed context bundle---containing sensor readings, timestamp, and a cryptographic signature tied to the device's hardware identity---can be presented on any RAT as evidence of operational continuity. Network-derived context (observed traffic patterns, peer behaviour analysis) is inherently non-portable.

\subsubsection{Network Trust: Fundamentally Non-Portable}

Network trust cannot be ported. The encryption, integrity, and confidentiality guarantees of a 5G link provide no assurance on a LoRaWAN channel. This represents the irreducible minimum cost of multi-RAT ZTA: at every cross-family transition, network trust must be re-established from the properties of the new RAT alone.

\subsubsection{The Proprietary Protocol Problem}

OcuSync and other proprietary protocols present a unique trust portability challenge: their closed nature prevents both the \textit{export} and \textit{import} of trust evidence. An OcuSync-authenticated session cannot produce a verifiable artefact consumable by external systems, and external trust evidence cannot be presented to OcuSync's internal authentication mechanism. This creates what we term a \textit{trust silo}---a domain from which no trust evidence can escape and into which no external trust evidence can enter. For multi-RAT UAVs that rely on OcuSync for command and control while using open protocols for data, this means the C2 trust domain and the data trust domain are architecturally isolated, with no possibility of unified trust management.

\subsection{Trust Composition for Parallel RAT Operation}
\label{sec:parallel}

The model thus far considers sequential RAT transitions. However, multi-RAT UAVs frequently operate on \textit{multiple RATs simultaneously}: OcuSync for C2, cellular for data, MAVLink for telemetry, and BLE for sensor interaction, all active concurrently.

\begin{definition}[Parallel Trust Composition]
\label{def:parallel}
When device $d$ is simultaneously connected to RATs $\{R_1, R_2, \ldots, R_m\}$ at time $t$, the composite trust score is:
\[
\mathcal{T}_{\parallel}(d, t) = \sum_{j \in \mathcal{J}} w_j \cdot f_j\bigl(s_j(d, t, R_1), \ldots, s_j(d, t, R_m)\bigr)
\]
where $f_j$ is a component-specific aggregation function.
\end{definition}

The choice of $f_j$ depends on the trust semantics. For \textbf{identity trust}, $f_{\mathrm{id}} = \max$: the device's identity is as trustworthy as its strongest authentication. For \textbf{device trust}, $f_{\mathrm{dev}} = \max$: device integrity is a property of the device, not the channel. For \textbf{network trust}, $f_{\mathrm{net}} = \min$: data is only as secure as the weakest channel actively carrying that data flow. For \textbf{contextual trust}, $f_{\mathrm{ctx}} = \mathrm{mean}$: contextual signals from multiple RATs can corroborate each other. For \textbf{policy trust}, $f_{\mathrm{pol}} = \min$: policy enforcement is only as strong as the weakest enforcement point.

The per-flow refinement for network trust is particularly important: a drone may transmit mission-critical telemetry over 5G (high network trust) while simultaneously broadcasting non-sensitive status over Meshtastic (low network trust). A single aggregate network trust score would underrepresent the actual security posture. We propose that mature implementations should compute per-flow trust scores, while the aggregate model provides a useful first approximation.

\subsection{Trust Portability Cost--Benefit Summary}

Table~\ref{tab:portability-cost} presents indicative cost comparisons for a representative RAT transition (5G $\rightarrow$ LoRaWAN) with and without trust portability mechanisms.

\begin{table*}[t]
\centering
\caption{Indicative Re-authentication Cost: Full vs.\ Trust-Portable (5G $\rightarrow$ LoRaWAN Transition)}
\label{tab:portability-cost}
\small
\begin{tabular}{@{}lcc@{}}
\toprule
\textbf{Approach} & \textbf{Energy (mJ)} & \textbf{Latency (ms)} \\
\midrule
Full ZT re-authentication on LoRaWAN & $\sim$850 & $\sim$6200 \\
Portable identity (HW attestation) + network re-auth only & $\sim$320 & $\sim$3800 \\
Portable identity + device + context & $\sim$180 & $\sim$2100 \\
\midrule
\textbf{Savings (maximum portability)} & \textbf{$\sim$79\%} & \textbf{$\sim$66\%} \\
\bottomrule
\end{tabular}
\medskip

\small\textit{Note}: Values are order-of-magnitude estimates based on published component power consumption figures for representative IoT hardware (ARM Cortex-M4 class processors, Semtech SX127x LoRa transceivers). Empirical validation on specific platforms is identified as future work.
\end{table*}

%=====================================================================
\section{Case Study: Multi-RAT UAV Mission}
\label{sec:casestudy}
%=====================================================================

We now apply the trust boundary model to a representative UAV mission, demonstrating the trust dynamics, power implications, and threat exposure of multi-RAT operation under ZTA.

\subsection{Mission Profile}

\textbf{Scenario}: A commercial quadrotor UAV conducts an agricultural infrastructure inspection in rural terrain over a 90-minute mission. The drone is a multi-RAT platform equipped with 5G/4G cellular (via eSIM), LoRaWAN (868\,MHz, for ground sensor data collection), Meshtastic (for opportunistic mesh relay), DJI OcuSync (for video and primary C2), and MAVLink over RFD900 (for telemetry to a ground monitoring station). The operator maintains a ZTA policy requiring $\mathcal{T} \geq \mathcal{T}_{\min} = 0.6$ for continued autonomous operation.

The mission proceeds through five phases:

\textbf{Phase~1} (0--10\,min): Launch from operations base. 5G connected via private cell. Full ZT authentication: eSIM-based 5G-AKA, TPM attestation, behavioural baseline established. OcuSync link active to operator controller. MAVLink telemetry streaming to GCS. $\mathcal{T} \approx 0.92$.

\textbf{Phase~2} (10--35\,min): Transit to survey area. 5G coverage lost at 2.1\,km; fallback to 4G LTE (3GPP handover, $\sigma_{\mathrm{id}} = 0.95$). 4G lost at 4.8\,km; data network switches to LoRaWAN via field gateway. Two trust boundary crossings. OcuSync and MAVLink links remain active throughout (direct radio, no network dependency). $\mathcal{T}$ drops from 0.92 to $\sim$0.71 (4G) then $\sim$0.48 (LoRaWAN)---below $\mathcal{T}_{\min}$.

\textbf{Phase~3} (35--65\,min): Survey operations. Data collection via LoRaWAN from ground sensors. Drone encounters neighbouring farm's Meshtastic mesh; policy engine evaluates opportunistic relay. One potential trust boundary crossing into an untrusted mesh domain. $\mathcal{T}$ remains low on data network; C2 trust via OcuSync is architecturally separate.

\textbf{Phase~4} (65--80\,min): Return transit. LoRaWAN $\rightarrow$ 4G LTE $\rightarrow$ 5G. Two trust boundary crossings with ascending trust recovery.

\textbf{Phase~5} (80--90\,min): Landing, data upload over 5G, mission debrief. Full trust restoration. $\mathcal{T} \approx 0.92$.

Total data-network trust boundary crossings: 5--6 in 90 minutes. Additionally, the OcuSync and MAVLink links operate as parallel, isolated trust domains throughout.

\subsection{Trust State Timeline}

\begin{figure*}[t]
\centering
\begin{tikzpicture}
\begin{axis}[
    width=\textwidth,
    height=5.5cm,
    xlabel={Mission Time (minutes)},
    ylabel={Composite Trust Score $\mathcal{T}$},
    xmin=0, xmax=90,
    ymin=0, ymax=1.0,
    grid=major,
    grid style={gray!30},
    legend pos=south east,
    legend style={font=\small},
    every axis label/.style={font=\small},
    every tick label/.style={font=\small},
]

% Trust threshold line
\addplot[dashed, trustlow, thick] coordinates {(0,0.6) (90,0.6)};
\addlegendentry{$\mathcal{T}_{\min} = 0.6$}

% Main trust score (data network)
\addplot[trusthigh, very thick] coordinates {
    (0,0.92) (10,0.90)
    (10,0.71) (12,0.78) (18,0.82)
    (18,0.35) (22,0.42) (28,0.48) (35,0.48)
    (35,0.22) (38,0.30) (42,0.35) (55,0.35) (65,0.35)
    (65,0.48) (68,0.55) (72,0.60)
    (72,0.71) (75,0.80) (78,0.85)
    (78,0.88) (82,0.90) (90,0.92)
};
\addlegendentry{Data network $\mathcal{T}$}

% OcuSync C2 trust (separate domain)
\addplot[trustrecov, thick, dotted] coordinates {
    (0,0.85) (10,0.84) (20,0.83) (30,0.82) (40,0.81)
    (50,0.80) (60,0.79) (70,0.78) (80,0.80) (90,0.85)
};
\addlegendentry{OcuSync C2 $\mathcal{T}$}

% Transition markers
\draw[trustlow, thick, ->] (axis cs:10,0.95) -- (axis cs:10,0.75) node[above right, font=\tiny] {5G$\rightarrow$4G};
\draw[trustlow, thick, ->] (axis cs:18,0.87) -- (axis cs:18,0.40) node[above right, font=\tiny] {4G$\rightarrow$LoRa};
\draw[trustlow, thick, ->] (axis cs:35,0.53) -- (axis cs:35,0.27) node[above right, font=\tiny] {LoRa$\rightarrow$Mesh};
\draw[trustrecov, thick, ->] (axis cs:65,0.30) -- (axis cs:65,0.43) node[below right, font=\tiny] {Mesh$\rightarrow$LoRa};
\draw[trustrecov, thick, ->] (axis cs:72,0.55) -- (axis cs:72,0.66) node[below right, font=\tiny] {LoRa$\rightarrow$4G};
\draw[trustrecov, thick, ->] (axis cs:78,0.82) -- (axis cs:78,0.85) node[below right, font=\tiny] {4G$\rightarrow$5G};

\end{axis}
\end{tikzpicture}
\caption{Trust score timeline for a 90-minute multi-RAT UAV mission. The data network trust score (solid blue) exhibits a sawtooth pattern with sharp drops at each RAT transition. The OcuSync C2 link (dotted green) operates as an isolated trust domain with slow natural decay. The drone spends approximately 48 minutes (53\%) of the mission with data-network trust below $\mathcal{T}_{\min} = 0.6$.}
\label{fig:sawtooth}
\end{figure*}
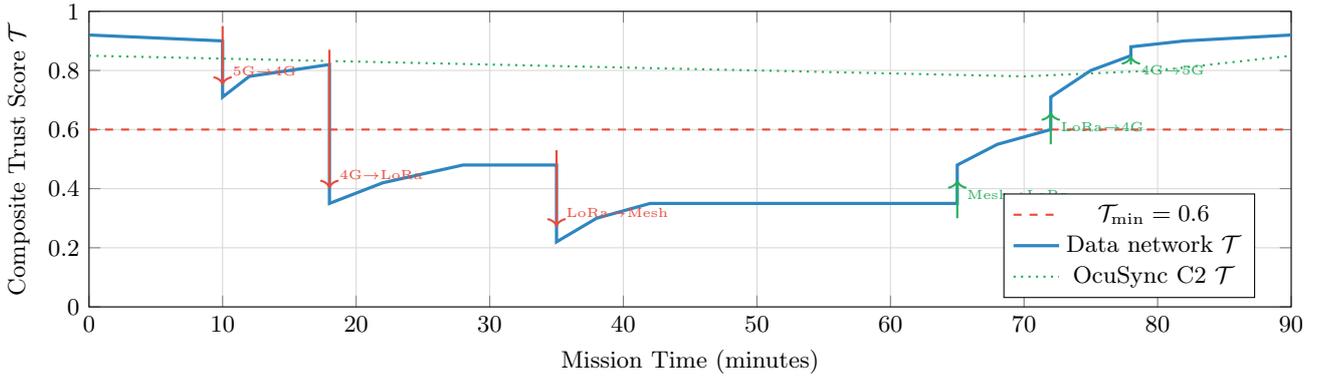

Figure~\ref{fig:sawtooth} illustrates the composite trust score across the mission duration for the data network. The characteristic \textit{sawtooth} pattern emerges: trust is high at launch, drops sharply at each RAT transition, partially recovers after re-authentication on the new RAT, and drops again at the next transition.

Three critical observations emerge:

\textbf{Sub-threshold exposure}: The drone operates below the minimum trust threshold for approximately 53\% of mission time on the data network. During this period, a strict ZTA implementation would either deny data network access (degrading mission capability) or accept elevated risk.

\textbf{Trust domain isolation}: The OcuSync C2 link maintains relatively stable trust throughout the mission because it operates on a direct radio link independent of the data network RAT transitions. However, this trust is opaque---the ZTA policy engine cannot verify OcuSync's internal trust state because the protocol is proprietary. The two trust domains (data network and C2) are architecturally decoupled, making unified trust assessment impossible.

\textbf{Asymmetric recovery}: Trust recovery after ascending transitions (LoRaWAN $\rightarrow$ 4G $\rightarrow$ 5G) is faster than the trust loss during descending transitions (5G $\rightarrow$ 4G $\rightarrow$ LoRaWAN) because higher-capability RATs can verify trust evidence more efficiently and support richer policy enforcement.

\subsection{Power Budget Analysis}

Table~\ref{tab:power-mission} breaks down the authentication energy cost for each RAT transition in the mission, comparing na\"ive full re-authentication against trust-portable approaches.

\begin{table*}[t]
\centering
\caption{Authentication Energy Cost for Multi-RAT UAV Mission}
\label{tab:power-mission}
\small
\begin{tabular}{@{}lcccl@{}}
\toprule
\textbf{Transition} & \textbf{Na\"ive (mJ)} & \textbf{Portable (mJ)} & \textbf{Saving} & \textbf{Notes} \\
\midrule
5G initial auth & 420 & 420 & --- & Baseline \\
5G $\rightarrow$ 4G & 280 & 95 & 66\% & 3GPP handover \\
4G $\rightarrow$ LoRaWAN & 850 & 180 & 79\% & Cross-family \\
LoRaWAN $\rightarrow$ Mesh & 120 & 45 & 63\% & PSK only \\
Mesh $\rightarrow$ LoRaWAN & 850 & 210 & 75\% & Cross-family \\
LoRaWAN $\rightarrow$ 4G & 280 & 110 & 61\% & Cross-family \\
4G $\rightarrow$ 5G & 180 & 60 & 67\% & 3GPP handover \\
\midrule
\textbf{Total} & \textbf{2980} & \textbf{1120} & \textbf{62\%} & \\
\bottomrule
\end{tabular}
\end{table*}

In isolation, the total authentication cost (2.98\,J na\"ive, 1.12\,J portable) appears modest against a typical survey drone's 10--15\,Wh (36--54\,kJ) battery. However, this single-pass cost is misleading for several reasons.

First, authentication cost is concentrated in the communications and compute subsystem, which typically consumes 5--8\% of total power (1.8--4.3\,kJ for a 90-minute mission). Second, ZTA mandates \textit{continuous} verification, not merely one-time authentication at RAT transitions. At a 30-second continuous verification interval, the cumulative cost over 90 minutes is approximately 180$\times$ the per-verification cost. Third, verification on bandwidth-constrained RATs (LoRaWAN at SF12: 51 bytes per uplink) requires multiple message exchanges spread over seconds, during which the radio must remain active.

Under continuous verification at 30-second intervals on the data network, the cumulative authentication power consumption ranges from approximately 15\% (trust-portable) to 40\% (na\"ive) of the communications/compute power budget. This is operationally significant: it directly reduces the power available for payload operations, telemetry, and flight control margin.

\subsection{Threat Model During Transitions}

RAT transitions create specific adversarial opportunities that do not exist during stable single-RAT operation:

\subsubsection{Adversary Model}

We adopt a modified Dolev-Yao adversary model~\cite{dolevyao1983} augmented with physical-layer capabilities. The adversary $\mathcal{A}$ can intercept, inject, and replay messages on any RAT within radio range, consistent with standard Dolev-Yao assumptions. Additionally, $\mathcal{A}$ possesses spectrum-level capabilities: selective jamming of specific RATs, transmission of rogue base station / gateway / mesh beacon signals, and the ability to observe (but not decrypt, absent key compromise) encrypted transmissions. $\mathcal{A}$ cannot break standard cryptographic primitives (AES, ECDH, HMAC-SHA256) within operationally relevant timescales. Crucially, $\mathcal{A}$ can influence which RAT the device selects by manipulating the RF environment---an extension beyond classical Dolev-Yao that is essential for modelling multi-RAT trust dynamics.

\subsubsection{Transition-Specific Threats}

\textbf{Forced downgrade attacks}: An adversary jams the higher-capability RAT (e.g., 5G) to force the device onto a lower-security RAT ($\beta^C$ becomes $\beta^A$), then exploits the weaker authentication model. This is particularly relevant given demonstrated vulnerabilities in LoRa physical-layer security and Meshtastic's PSK-only model.

\textbf{Rogue RAT injection}: During the transition window, when the device is searching for a new RAT, an adversary presents a malicious base station, gateway, or mesh node. The device, operating under pressure to restore connectivity, may authenticate to the rogue infrastructure.

\textbf{Trust evidence replay}: Portable trust artefacts captured during one transition are replayed during a later transition to impersonate the device. Freshness guarantees (timestamps, nonces) mitigate this but add overhead.

\textbf{Transition window exploitation}: The interval between disconnection from $R_i$ and authenticated connection to $R_{i+1}$ is a ``trust gap'' during which the device operates with degraded or no trust. An adversary aware of this pattern can time attacks to coincide with predicted transition windows. The exploitation probability is strongly dependent on transition type: for planned transitions ($\beta^P$) with pre-staged authentication, the trust gap duration is minimal (estimated 50--200\,ms) and exploitation probability is low ($p_{\mathrm{exploit}} \approx 0.02$--$0.05$). For adversary-forced transitions ($\beta^A$), the trust gap extends to 2--15 seconds while the device searches for alternative RATs under adversarial RF conditions, increasing exploitation probability by a factor of 2--3$\times$ ($p_{\mathrm{exploit}} \approx 0.10$--$0.25$).

\textbf{C2/data domain divergence}: The architectural isolation between OcuSync C2 and the data network creates a scenario where an adversary compromises the data network trust domain without affecting C2 (or vice versa). The ZTA policy engine, unable to assess OcuSync's internal state, cannot detect cross-domain inconsistencies.

%=====================================================================
\section{Discussion}
\label{sec:discussion}
%=====================================================================

\subsection{Implications for ZTA Standards}

The NIST SP 800-207 reference architecture does not address RAT-level trust boundaries. The PE/PA/PEP model assumes continuous connectivity between the policy infrastructure and the device, and does not define behaviour when the device transitions to a network where the PE is unreachable or the PEP cannot enforce the current policy set.

We identify three extensions required for multi-RAT ZTA:

First, a \textbf{disconnected policy engine} capability, enabling devices to make local trust decisions based on cached policy when the PE is unreachable on the current RAT. This is analogous to the Denied, Disrupted, Intermittent, Limited (DDIL) network models studied in military contexts~\cite{saghaian2023bayesian,vanderlinden2023milcom}.

Second, \textbf{trust boundary crossing event handlers} as a formal component of the ZTA architecture, defining procedures for trust evidence export (before leaving a RAT), trust state assessment (during transition), and trust evidence import (after joining a new RAT).

Third, \textbf{RAT trust profiles} that characterise the trust capabilities and limitations of each RAT family, enabling the PE to adjust trust expectations and verification requirements based on the currently active RAT. A trust threshold of $\mathcal{T}_{\min} = 0.8$ may be achievable on 5G but physically impossible on Meshtastic; the policy framework must accommodate RAT-dependent trust ceilings.

\subsection{RAT-Dependent Trust Ceilings}

As the worked example (Section~\ref{sec:worked-example}) demonstrated, a device on LoRaWAN cannot achieve $\mathcal{T}_{\min} = 0.6$ regardless of authentication effort. We formalise this as a \textit{trust ceiling}:

\begin{definition}[Trust Ceiling]
The trust ceiling $\hat{\mathcal{T}}(R)$ for RAT $R$ is the maximum achievable composite trust score on $R$ given $R$'s trust infrastructure capabilities:
\[
\hat{\mathcal{T}}(R) = \sum_{j \in \mathcal{J}} w_j \cdot \hat{s}_j(R)
\]
where $\hat{s}_j(R)$ is the maximum achievable value of trust component $j$ on RAT $R$.
\end{definition}

When $\hat{\mathcal{T}}(R) < \mathcal{T}_{\min}$, the policy framework must either accept a RAT-dependent threshold ($\mathcal{T}_{\min}(R) \leq \hat{\mathcal{T}}(R)$), restrict operations on that RAT, or require compensating controls (e.g., increased monitoring via parallel high-trust RATs).

\subsection{Regulatory Compliance and Remote ID}

The Remote ID analysis (Section~\ref{sec:remoteid}) reveals a regulatory-security disconnect. Regulators mandate identity broadcasting for airspace safety, but the broadcast mechanism provides no ZTA-relevant trust assurance. A drone compliant with Remote ID is visible but not authenticated. UTM (Unmanned Traffic Management) systems should treat Remote ID as a \textit{contextual hint} requiring corroboration from authenticated sources, not as an identity assertion. Future Remote ID standards should consider incorporating lightweight message authentication to elevate Remote ID from a weak to moderate trust signal.

\subsection{The Case for Power-Aware Trust Metrics}

Our analysis reveals a fundamental tension: Zero Trust demands maximal verification, but multi-RAT IoT devices face non-trivial power costs for each verification event, amplified by the frequency of RAT transitions. This tension cannot be resolved by better authentication protocols alone---it requires a metric that jointly optimises trust level and energy cost.

Specifically, we require a metric $M$ such that:
\begin{equation}
M = \frac{f(\mathcal{T})}{g(P)}
\label{eq:metric-requirement}
\end{equation}
where $f(\mathcal{T})$ is an increasing function of composite trust and $g(P)$ is an increasing function of power consumed. Maximising $M$ under constraints $\mathcal{T} \geq \mathcal{T}_{\min}$ and $P \leq P_{\max}$ yields the trust-optimal, power-feasible verification strategy.

In companion work, we formalise this as the \textit{Trust-per-Watt} (TpW) optimisation framework, defining $\TpW = \mathcal{T}/P$ as a formally analysable metric with provable optimisation properties. The trust boundary model and trust survival functions presented in this paper provide the input parameters for TpW optimisation: the survival matrices define how much trust is available ``for free'' at each transition, and the recovery costs define the price of rebuilding what is lost.

\subsection{Defence and Multi-Domain Operations}

The multi-RAT trust boundary problem is acutely relevant to military UAV operations, where spectrum availability is contested and RAT switching may be adversary-induced rather than coverage-driven. Coalition operations compound the challenge: a UAV operating within a FVEY (Five Eyes) coalition must traverse trust domains belonging to different national military networks, each with its own identity infrastructure and classification policies. The trust portability problem at the RAT level mirrors the coalition trust federation problem studied by Van der Linden et al.~\cite{vanderlinden2023milcom}, but at a lower protocol layer.

\subsection{Commercial IoT Applications}

The trust boundary model applies beyond UAVs to any multi-RAT IoT deployment:

\textbf{Connected autonomous vehicles}: A self-driving vehicle transitions between 5G, C-V2X (Cellular Vehicle-to-Everything), DSRC (Dedicated Short-Range Communications), and satellite as it moves through urban, suburban, and rural environments. Each transition is a trust boundary crossing under UNECE WP.29 cybersecurity regulations.

\textbf{Supply chain tracking}: An asset tracker on a shipping container transitions from warehouse Wi-Fi to cellular to satellite at sea to LoRaWAN at the destination port. Continuous chain-of-custody trust requires trust portability across these transitions.

\textbf{Smart agriculture}: Ground sensors reporting via LoRaWAN interact with survey drones (multi-RAT), edge gateways (cellular + LPWAN), and cloud analytics (enterprise network). Each data handoff crosses a trust boundary.

\subsection{The Proprietary Protocol Challenge}

Our analysis identifies proprietary protocols---exemplified by DJI OcuSync---as a structural impediment to unified Zero Trust in multi-RAT environments. When a drone's command-and-control link operates on a closed protocol, the ZTA policy engine cannot assess, verify, or influence the trust state of that link. This creates a ``shadow trust domain'' operating outside ZTA governance.

Three potential approaches exist: vendor API integration (DJI or other vendors expose trust-relevant signals via an authenticated API), protocol reverse engineering (security researchers analyse the protocol to assess its trust properties, as demonstrated by Schiller et al.~\cite{schiller2023dronesec}), or architectural isolation (accept that the C2 domain is outside ZTA governance and design compensating controls). None is fully satisfactory, and this remains an open problem for the drone security community.

\subsection{Limitations}

This paper presents a qualitative framework and the trust survival values ($\sigma_j$) are reasoned assessments based on protocol analysis rather than empirical measurements on specific hardware. The power cost estimates are indicative, derived from published component specifications rather than end-to-end measured authentication sequences. The trust state model assumes independence between trust components, which may not hold in practice. In reality, trust components exhibit positive correlations: a compromised device identity ($s_{\mathrm{id}} \rightarrow 0$) implies potential device compromise ($s_{\mathrm{dev}}$ should decrease), and loss of network trust ($s_{\mathrm{net}} \rightarrow 0$) degrades the PE's ability to enforce policy ($s_{\mathrm{pol}}$ decreases). A more complete model would incorporate an inter-component correlation matrix $\boldsymbol{\Sigma} \in [0,1]^{|\mathcal{J}| \times |\mathcal{J}|}$ such that a change in component $j$ propagates to component $k$ proportional to $\Sigma_{jk}$. For example, with identity-device correlation $\Sigma_{\mathrm{id,dev}} = 0.6$, a complete identity trust loss would induce a 60\% reduction in device trust even absent direct evidence of device compromise. Future work will formalise these inter-component dependencies via covariance terms in the trust state model, potentially drawing on Bayesian network representations to capture conditional dependencies. The exponential trust decay model is a first-order approximation; as noted in Section~\ref{sec:decay}, Weibull or event-triggered decay profiles may better capture specific operational dynamics. The analysis considers single-device scenarios; swarm-level trust management, where devices can vouch for one another, introduces additional complexity.

These limitations are deliberate: the contribution of this paper is the \textit{conceptual framework}---trust boundary crossings, trust survival functions, trust portability---rather than empirical validation. The framework establishes the vocabulary and analytical tools for subsequent empirical and formal work.

%=====================================================================
\section{Future Work}
\label{sec:future}
%=====================================================================

This paper opens several research directions:

\textbf{Trust-per-Watt formalisation}: The companion paper formalises TpW as the optimisation objective for the trust boundary crossing problem, proving properties of the resulting framework including convexity, complexity, and approximation guarantees.

\textbf{Formal verification}: Model trust boundary crossings in the applied $\pi$-calculus or a suitable process algebra. Verify that trust portability mechanisms preserve security invariants across RAT transitions. Prove trust floor guarantees under adversarial RAT manipulation.

\textbf{Empirical validation}: Construct a multi-RAT testbed incorporating LoRa (Semtech SX127x), cellular (Quectel 5G/4G modem), Meshtastic, and MAVLink (RFD900) interfaces on a common platform (e.g., Raspberry Pi or STM32). Measure actual authentication costs, trust recovery latencies, and power consumption for each RAT transition.

\textbf{Game-theoretic RAT selection}: Model adversarial RAT switching as a two-player game between a defender (which RAT to select, how much to re-authenticate) and an attacker (which RAT to jam, spoof, or exploit). Compute Nash equilibrium strategies and characterise the price of anarchy for sub-optimal trust management.

\textbf{Swarm trust delegation}: Extend to multi-device scenarios where trust evidence can be delegated, vouched for, or collectively attested by swarm members, reducing per-device re-authentication cost at the expense of trust chain depth.

\textbf{Remote ID authentication}: Design and evaluate lightweight Remote ID message authentication mechanisms compatible with ASTM F3411 broadcast constraints, elevating Remote ID from a weak contextual signal to a moderate trust input.

\textbf{6G and emerging RAT integration}: The 3GPP Release 19/20 roadmap for 6G introduces native support for non-terrestrial networks, sub-THz communications, AI-native air interfaces, and integrated sensing and communications (ISAC). These capabilities will expand the multi-RAT landscape further and introduce new trust-relevant properties---particularly AI-driven network management that may dynamically alter trust characteristics. The trust boundary model should be extended to accommodate 6G-specific trust signals, including AI-verifiable network behaviour attestation and sub-THz channel authentication via physical-layer security.

\textbf{Empirical decay rate calibration}: Measure trust decay rates $\lambda_j(R)$ on specific hardware platforms across all eight RAT families, validating or refining the exponential decay model.

%=====================================================================
\section{Conclusion}
\label{sec:conclusion}
%=====================================================================

Zero Trust Architecture and multi-RAT IoT are two of the most significant trends in network security and wireless communications respectively. Their intersection---where mobile devices must maintain continuous trust verification while dynamically switching between fundamentally different radio technologies---has not been systematically studied.

We have shown that each RAT transition constitutes a trust boundary crossing that invalidates some or all trust state, requiring costly re-establishment. The trust survival analysis across seven RAT families---including both open-standard and proprietary drone communication protocols---reveals that cross-family identity and network trust survival is near-zero, making re-authentication the dominant cost. Device trust, anchored in hardware roots of trust, exhibits the highest cross-RAT portability.

Trust portability mechanisms can reduce re-authentication costs by 60--80\% for amenable trust components, but are fundamentally blocked by proprietary protocols (OcuSync) that form impenetrable trust silos, and by the inherent non-portability of network trust. For power-constrained mobile IoT devices such as UAVs, the cumulative cost of Zero Trust compliance across multi-RAT missions creates a tension that demands power-aware trust optimisation.

The trust boundary model, trust survival functions, and trust portability framework presented here provide the conceptual foundation and analytical vocabulary for addressing this challenge. In companion work, we develop Trust-per-Watt as the formal optimisation metric for this problem, completing the bridge from problem identification to solution framework.

%=====================================================================
% Acknowledgements
%=====================================================================

\section*{Acknowledgements}

The author gratefully acknowledges the supervision of Professor Andrew Martin (Department of Computer Science, University of Oxford). This work was conducted as part of a DPhil research programme at the University of Oxford.

%=====================================================================
% References
%=====================================================================

\end{document}